\documentclass{moriond}

\def\be{\begin{equation}}
\def\ee{\end{equation}}
\def\bea{\begin{eqnarray}}
\def\eea{\end{eqnarray}}

\usepackage{amsmath}

\newcommand{\Photo}{}

\begin{document}
\vspace*{4cm}
\title{SHEDDING NEW LIGHT ON THE $\bf{B \to \pi K}$ PUZZLE}

\author{E. MALAMI}

\address{DAMTP, University of Cambridge, Centre for Mathematical Sciences, Wilberforce Road, CB3 0WA, Cambridge, United Kingdom}

\maketitle\abstracts{The $B \to \pi K$ system provides a rich laboratory for testing the Standard Model and studying CP violation. A particularly important channel is $B^0_d\to\pi^0 K_{\rm S}$, the only mode exhibiting both direct and mixing-induced CP violation. Recent Belle II measurements of the CP asymmetries in this decay provide valuable new input. An updated analysis incorporating these new data provides new insight on the long-standing $B \to \pi K$ puzzle. Looking ahead to the high-precision era of flavour physics, the  $B \to \pi K$ system can be further exploited to potentially reveal new sources of CP violation.}

\section{Introduction}
The $B \to \pi K$ decays serve as a key laboratory for testing the Standard Model (SM) of flavour physics and for probing CP-violating effects. The system consists of four hadronic decay channels: $B^+\to\pi^0K^+$, $B^+\to\pi^+K^0$,$B^0_d\to \pi^-K^+$ and $B_d^0 \to \pi^0 K^0$. Inconsistencies arise among the branching ratios and the CP asymmetries in these four channels, resulting in a puzzling situation. Tree-level contributions are strongly Cabibbo-suppressed due to the smallness of the CKM matrix element $|V_{ub}|$. As a result, these decays are dominated by QCD (gluonic) penguin amplitudes, while electroweak penguin (EWP) contributions also play an important role. In particular, potential New Physics (NP) effects may enter via EWPs.

Among the four channels, the most prominent channel is the $B^0_d\to\pi^0 K_{\rm S}$, as it is the only mode exhibiting both direct and mixing-induced CP violation. This makes it particularly interesting for precision studies of CP violation, especially in the context of time-dependent (mixing-induced) asymmetries. The $B \to \pi K$ system has been extensively studied in the literature (see, for instance,~\cite{NQ,GHLR,GRL,RF-96,BF-98,Neu-98,BeNe,FRS,groro,BGV,BFRS-2}). A renewed motivation for revisiting this system comes from recent Belle II results on CP asymmetries in $B_d^0 \to \pi^0 K_S$ \cite{Veronesi:2023dak}. Using as a basis our previous analysis in \cite{Fleischer:2018bld}, we provide an updated study \cite{tbw} aiming to shed light on the following questions: what is the picture arising from the current data? Are there really inconsistencies among the branching ratios and CP asymmetries of these decays?

\section{Setting the Stage}

\subsection{Hadronic Parameters}
The $B \to \pi K$ decays, as non-leptonic modes, are challenging due to the presence of hadronic matrix elements of four-quark operators entering the corresponding low-energy effective Hamiltonians. Flavour symmetries of strong interactions imply relations between the $B \to \pi K$ amplitudes and those of the $B \to \pi\pi$ and $B \to KK$ systems, allowing one either to reduce the hadronic uncertainties or to determine them from experimental information on the latter decay modes. In our analysis, we aim to keep theoretical assumptions about strong interactions as minimal as possible and include SU(3)-breaking effects using QCD factorisation \cite{BeNe}. To make this explicit, we introduce and determine the hadronic parameters, thereby providing a full update of the analysis based on the most recent experimental information.

The tree and QCD penguin contributions are parametrised as
\bea \label{eq:hadronic_par}
    r e^{i\delta} \equiv (\hat{T} - \hat{P}_{tu})/P', \qquad
    r_{\rm c} e^{i\delta_{\rm c}} \equiv (\hat{T} + \hat{C})/P'.
\eea
Here $\hat{T}$ and $\hat{C}$ denote the colour-allowed and colour-suppressed tree amplitudes, respectively. The quantities $P_{tc}$ and $P_{tu}$ describe QCD penguin contributions with internal $(t,c)$ and $(t,u)$ quarks, respectively, while $P'$ is proportional to the QCD penguin amplitude $P_{tc}$. Using SU(3) flavour symmetry, these parameters are related to those of the $B \to \pi\pi$ system, which is used as input for their determination. We allow for non-factorisable SU(3)-breaking effects at the level of $20\%$ and obtain the following updated values based on the most recent $B \to \pi\pi$ data~\cite{tbw}:
\bea \label{eq:hadronic_par_values}
    r=0.10 \pm 0.02,  \qquad \delta = (31.4 \pm 20.4)^o \\
    r_{\rm c}= 0.17 \pm 0.03, \qquad \delta_{\rm c} = (0.68 \pm 20.6)^o .
\eea

The colour-suppressed EWP contributions are described by the hadronic parameters $\rho_{\rm c}$ and $\theta_{\rm c}$, which are determined using U-spin symmetry and input from the $B \to KK$ system. This yields the updated values~\cite{tbw}:
\bea \label{eq:hadronic_par_values2}
    \rho_{\rm c}=0.02 \pm 0.01,  \qquad \theta_{\rm c} = (1.7 \pm 6.1)^o .
\eea

These hadronic inputs will serve as the basis for the numerical analysis of CP asymmetries and their correlations in the $B \to \pi K$ system presented below.

\subsection{Electroweak Penguin Parameters ${q}$ and ${\phi}$}
The EWP contributions in the $B \to \pi K$ system are described by two parameters; $q$, which measures the strength of EWP topologies with respect to tree amplitudes, and $\phi$, which denotes a CP-violating phase. They are defined as \cite{RF-95,NR}
\bea\label{eq:q_phi}
q e^{i\phi} e^{i\omega} \equiv - \left(\frac{\hat{P}_{EW} + \hat{P}_{EW}^{\rm C}}{\hat{T} +\hat{C}}\right) {=} \frac{-3}{2\lambda^2 R_b}\left(\frac{C_9 + C_{10}}{C_1 + C_2}\right) R_q = (0.69 \pm 0.3) R_q,
\eea
where $\hat{P}_{EW}$ and $\hat{P}_{EW}^{\rm C}$ denote the colour-allowed and colour-suppressed electroweak penguin contributions, respectively. The phase $\omega$ is CP-conserving and vanishes in the SU(3) limit. The parameter $\lambda = |V_{us}|$ is the Wolfenstein parameter of the CKM matrix, while $R_b$ denotes one of the sides of the unitarity triangle. The quantities $C_i$ are Wilson coefficients of the low-energy effective weak Hamiltonian. The parameter $R_q = 1.00 \pm 0.30$ encodes SU(3)-breaking corrections. In the SM, $\phi = 0^\circ$, and the parameter $q$ can be determined from Eq.~\ref{eq:q_phi}.

\section{Correlation between the CP Asymmetries}
Using the framework introduced above, in particular the electroweak penguin parameters $q$ and $\phi$, we now turn to the correlations between CP asymmetries in the $B \to \pi K$ system. Throughout this section, we work within the SM framework, where $q = 0.69$ and $\phi$ vanishes.

The time-dependent~CP~asymmetry for neutral $B_d^0$ decays into a CP eigenstate $f$ is:
\bea \label{eqCP-Asy-t}
\mathcal{A}_\text{CP}(t) \equiv \frac{\Gamma(\bar{B}_d^0(t) \rightarrow f) - \Gamma(B_d^0(t) \rightarrow f)}{\Gamma(\bar{B}_d^0(t) \rightarrow f) + \Gamma(B_d^0(t) \rightarrow f)} 
= A_\text{CP}^f \cos(\Delta M_d t) + S_\text{CP}^f \sin(\Delta M_d t)\ .
\eea
The time dependence originates from $B_d^0$–$\bar{B}_d^0$ oscillations, with $\Delta M_d \equiv M^{(d)}_{\rm H} - M^{(d)}_{\rm L}$ denoting the mass difference between the heavy and light mass eigenstates of the $B_d$ system \cite{Fle02}. The coefficient $A_\text{CP}^f$ characterises direct CP violation, while $S_\text{CP}^f$ describes mixing-induced CP violation.

\begin{figure}[t!]
\centerline{\includegraphics[width=0.45\linewidth]{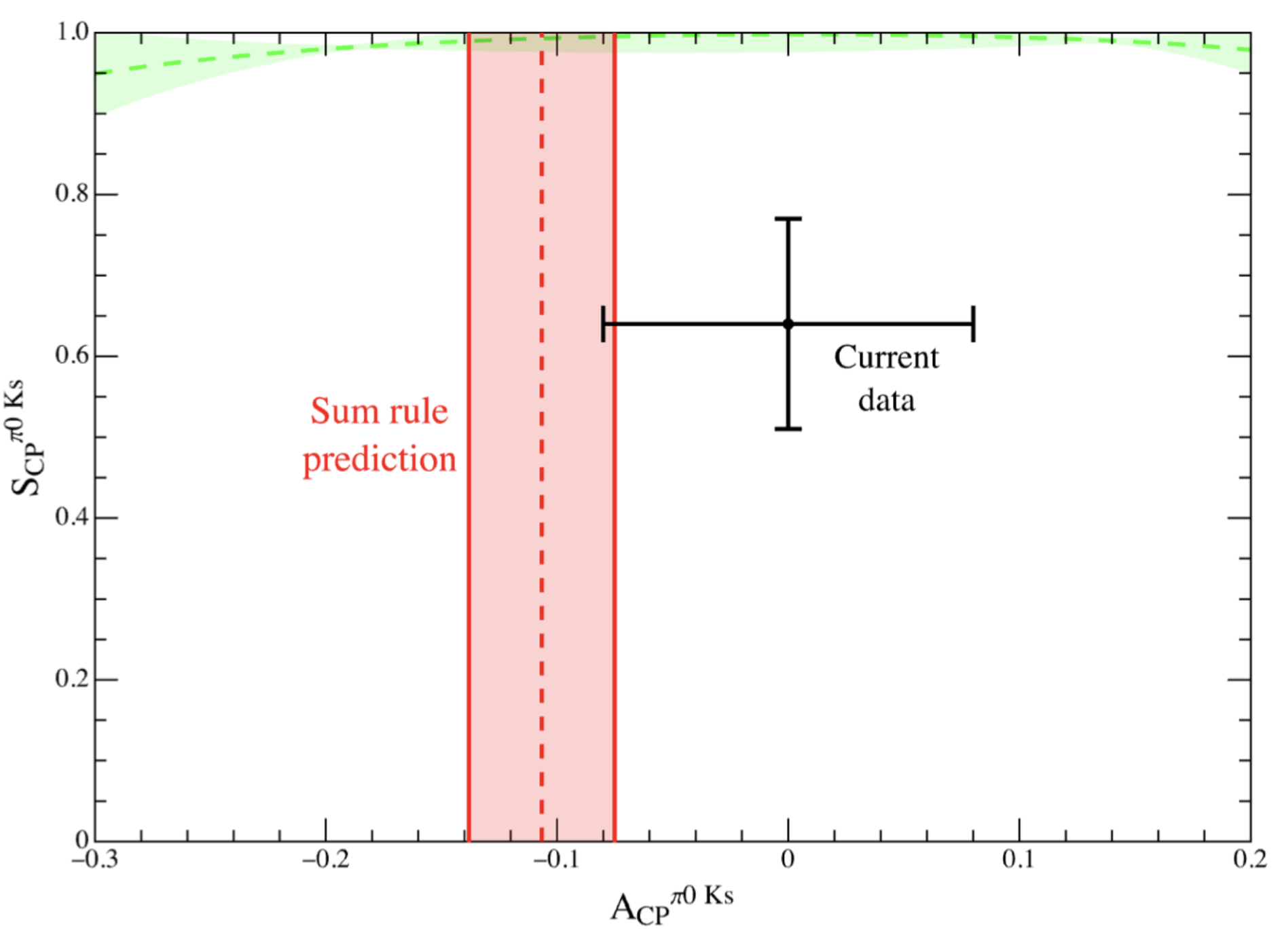}}
\caption[]{Correlations between $S_{\rm CP}^{\pi^0 K_S}$ and $A_{\rm CP}^{\pi^0 K_S}$ taking the uncertainties into account~\cite{tbw}.}
\label{fig:correlations}
\end{figure}

The amplitudes of the $B \to \pi K$ system satisfy the isospin relation \cite{NQ,GHLR}
\bea
\sqrt{2} A(B_d^0\to \pi^0 K^0) +
A(B^0\to \pi^- K^+) = - (\hat{T}' +\hat{C}')e^{i\gamma} + \left(\hat P'_{EW}+{\hat{ P}^{'{\rm C}}_{EW}}\right) \equiv 3A_{3/2} ,
\eea
with an analogous relation for the CP-conjugate amplitudes. Here, $A_{3/2}$ denotes the isospin $I=3/2$ amplitude, and $\gamma$ is the UT angle, for which we use $\gamma = (66.4 \pm 2.8)^\circ$ \cite{HeavyFlavorAveragingGroupHFLAV:2024ctg}. This relation provides the basis for a geometrical construction of isospin triangles, from which CP-violating observables can be related, with only minimal SU(3) input.

Following the strategy outlined in \cite{Fleischer:2018bld}, the corresponding isospin triangles can be constructed in the complex plane, allowing us to derive correlations between CP asymmetries. This construction allows us to express the mixing-induced CP violation in terms of the underlying amplitude phases. In particular, the mixing-induced CP asymmetry of the $B_d^0 \to \pi^0 K_S$ mode is given by~\cite{FJPZ}
\bea
S_{\rm CP}^{\pi^0K_{\rm S}}  = \sqrt{1- (A^{\pi^0K_{\rm S}}_{\rm CP})^2} \sin(\phi_d -\phi_{00}), 
\eea
where $A^{\pi^0K_{\rm S}}_{\rm CP}$ denotes the direct CP asymmetry, $\phi_d$ is the $B_d^0$–$\bar{B}_d^0$ mixing phase, and $\phi_{00}$ is the relative phase between the decay amplitudes of $B_d^0 \to \pi^0 K^0$ and its CP-conjugate process. The isospin construction leads to a four-fold ambiguity in $\phi_{00}$, corresponding to four possible orientations of the amplitude triangles, and hence to four distinct contours in the $S_{\rm CP}^{\pi^0 K_S}$–$A_{\rm CP}^{\pi^0 K_S}$ plane. This ambiguity can be resolved by employing additional information on the hadronic parameters $r_{\rm c}$ and $\delta_{\rm c}$.

The resulting contour, which resolves the four-fold ambiguity, is shown as the green band in Fig.~\ref{fig:correlations} and represents the theoretical prediction in the $S_{\rm CP}^{\pi^0 K_S}$–$A_{\rm CP}^{\pi^0 K_S}$ plane. Compared to our previous analysis, the updated input leads to a slight upward shift of this band. The current world averages, indicated by the black point in Fig.~\ref{fig:correlations}, are given by \cite{ParticleDataGroup:2024cfk}:
\bea
A_{\text{CP}}({B_d \to \pi^0 K^0})= 0.00 \pm 0.08, \\
S_{\text{CP}}({B_d \to \pi^0 K^0})= 0.64 \pm 0.13 .
 \label{eq:S,A}
\eea
Comparing the experimental averages with the theoretical prediction (green band), we find that the so-called $B \to \pi K$ puzzle persists. 

A particularly interesting test of the SM is provided by the sum rule \cite{Gronau:2005kz,Gronau:2006xu}
\begin{eqnarray}
\label{eq:sum-rule-I}
\Delta_{\rm SR}^{({\rm I})} &=& A_\text{CP}^{\pi^\pm K^\mp} +  A_\text{CP}^{\pi^\pm K^0} 
\frac{\mathcal{B} r(B^{+}\to\pi^{+}  K^0)}{\mathcal{B} r (B^0_d\to\pi^-  K^+)} \frac{\tau_{B^0}}{\tau_{B^+}} 
- A_\text{CP}^{\pi^0K^\pm} \frac{2 {\mathcal{B} r(B^{+}\to\pi^0  K^+)}}{\mathcal{B}r(B^0_d\to\pi^-  K^+)} 
\frac{\tau_{B^0}}{\tau_{B^+}} \nonumber \\ 
&-& A_\text{CP}^{\pi^0 K^0} \frac{2  \mathcal{B}r(B^0_d\to\pi^0  K^0)}{\mathcal{B}r(B^0_d\to\pi^-  K^+)} 
= 0 + {\cal O}(r_{({\rm c})}^2,\rho_{\rm c}^2) \ ,
\end{eqnarray}
which uses only information from CP-averaged branching ratios and CP asymmetries, with experimental inputs taken from PDG \cite{ParticleDataGroup:2024cfk}. We observe that the sum-rule band (denoted in red in Fig.~\ref{fig:correlations}) has shifted slightly rightwards, moving closer to the current world-average point, with the corresponding uncertainty regions now showing mild overlap. \\

It is important to further explore this system, in particular in view of possible NP effects. In this context, it is instructive to study how the picture changes when deviating from the SM values. The Belle II experiment is now playing a key role in studies of $B^0_d\to\pi^0 K_{\rm S}$, opening new opportunities for precision measurements that will be further complemented by the upcoming LHCb upgrade. These developments mark an exciting era for flavour physics.

\section*{Acknowledgments}

I would like to thank Robert Fleischer, Shiyang Ming, Keri Vos, Chenxi Zhang, and Ya\v{g}mur Zubaro\v{g}lu for the collaboration throughout this analysis and for their contributions to the results presented in this work, as well as Ben Allanach for useful discussions. I am also grateful to the organisers of Moriond for the opportunity to present this work. This research has been partially supported by the Netherlands Organisation for Scientific Research (NWO) and the STFC consolidated grant ST/X000664/1.

\section*{References}


\end{document}